\documentclass[modern,medium]{oup-authoring-template}

\onecolumn

\theoremstyle{thmstyletwo}%

\newtheorem{thm}{Theorem}[section]

\newtheorem{cor}{Corollary}[section]

\newtheorem{hyp}{Hypothesis}

\newcommand{\R}{\mathbb{R}}
\newcommand{\Z}{\mathbb{Z}}
\newcommand{\T}{\mathbb{T}}
\newcommand{\M}{\mathsf{M}}
\newcommand{\MS}{\mathsf{S}}
\newcommand{\K}{\mathsf{K}}
\newcommand{\I}{\mathsf{I}}
\newcommand{\metric}{\ensuremath{\mathrm{g}}}

\newcommand{\submetrict}{\ensuremath{\gamma}}

\newcommand{\grad}{\raisebox{0.01em}{\scalebox{1.1}[1.4]{${\scriptstyle \nabla}$}}}

\newcommand{\gradD}{\raisebox{0.01em}{\scalebox{1.2}[1.0]{${\mathrm{D}}$}}}
\newcommand{\energy}{\ensuremath{\mathrm{\rho}}}
\newcommand{\energyt}{\ensuremath{\mathsf{Q}_{E}}}
\newcommand{\energytt}{\ensuremath{\mathrm{\mu}}}
\newcommand{\energyttt}{\ensuremath{\mathrm{\rho}_E}}

\newcommand{\pressure}{\scalebox{0.9}[1.0]{\ensuremath{\mathsf{P}}}}

\newcommand{\hubble}{\ensuremath{\mathrm{H}}}
\newcommand{\hubblet}{\ensuremath{\mathrm{U}}}
\newcommand{\hubblev}{\ensuremath{\mathrm{V}}}

\newcommand{\eqstate}{\ensuremath{\mathrm{w}}}

\begin{document}

\journaltitle{$*********$}
\DOI{$*********$}
\copyrightyear{2024}
\pubyear{2024}
\access{Submitted: 02-Aug-2024}
\appnotes{Paper}

\firstpage{1}


\title[Einstein's equations constrained by homogeneous and isotropic expansion]{Einstein's equations constrained by homogeneous and isotropic expansion: Initial value problems and applications}

\author[1,$\ast$]{Leandro G. Gomes\ORCID{0000-0002-3461-3206}}
\author[1]{Marcelo A. C. Nogueira}
\author[1]{Lucas Ruiz dos Santos}

\authormark{L.G. Gomes,  M. A. C. Nogueira, L. Ruiz dos Santos}

\address[1]{\orgdiv{Department of Mathematics}, \orgname{Federal University of Itajuba}, \orgaddress{\street{Av. BPS, 1303}, \postcode{37500-903}, \state{MG}, \country{Brazil}}}

\corresp[$\ast$]{Corresponding author. \href{email:lggomes@unifei.edu.br}{lggomes@unifei.edu.br}}

\received{Date}{0}{Year}
\revised{Date}{0}{Year}
\accepted{Date}{0}{Year}

\abstract{In this manuscript, we put forth a general scheme for defining initial value problems from Einstein's equations of General Relativity constrained by homogeneous and isotropic expansion. The cosmological models arising as solutions are naturally interpreted as spatially homogeneous and isotropic on ``large scales". In order to show the well-posedness and applicability of such a scheme, we specialize in a class of spacetimes filled with the general homogeneous perfect fluid and inhomogeneous viscoelastic matter. We prove the existence, uniqueness, and relative stability of solutions, and an additional inequality for the energy density. As a consequence of our theorems, a new mechanism of energy transfer appears involving the different components of matter. A class of exact solutions is also obtained to exemplify the general results. 
}


\maketitle

\section{Introduction}\label{sec:Intro} 
%

There are two main hypotheses supporting the standard model of Cosmology, namely, that the universe is homogeneous on large scales and the cosmic expansion is isotropic \cite{CQG_2010_Maartens_Clarkson, 2017_MNRAS_Bernui}. However, the standard approach to Cosmology, which is built over the Friedmann-Lemaître-Robertson-Walker (FLRW) spacetimes \cite{Book_1993_Peebles}, has been questioned and put under intense scrutiny in recent years \cite{JHEA_2022_HubbleTensionReview, ObservationalCosmologicalPrinciple}. All those uncertainties in our current understanding of the Universe we live in bring about some sort of problems that, at the ultimate level, pertain to the mathematical foundations of the theory. The formal understanding of the basic constituents of the cosmological principle \cite{LGGomes_2024_CQG_1} and the precise description of the concept of homogeneous and isotropic expansion \cite{LGGomes_2024_PeriodicHomIsotExpansion} are two recent examples of the need for a more robust understanding of the basic assumptions largely used by the working cosmologist. In this context, here we aim to set (and solve) a class of initial value problems arising from Einstein's equations which, as a two-way connection, is equally interesting in the area of Mathematical General Relativity and their prospects of applications in Cosmology.


We start with the product manifold $\M=\I \times \MS$, $\I \subset \R$ an open interval and $\MS$ a oriented manifold of dimension $\dim\M=m-1$, the time function $t: \M \to \R$ given by the projection $(t,x)\mapsto t $, and the Lorentzian metric  
\begin{equation}\label{Eq:MetricGeneral}
 \metric = - \, e^{2 \phi(t,x)} dt^2 + a(t,x)^2\, \submetrict (t) \, ,
\end{equation}
where $\submetrict(t_1)$ is a Riemannian metric in $\MS \cong \{t_1\}\times\MS$ for each $t_1 \in \I$ and $a:\M \to \R$ a scale factor, uniquely determined up to a multiplicative function $a_0:\MS \to \R$ by demanding the volume $(m-1)$-form of $\submetrict(t)$ to be time-invariant (see Ref. \cite{LGGomes_2024_PeriodicHomIsotExpansion}). In this case, the Hubble parameter, or the mean curvature of the second fundamental form (expansion tensor), is determined  through
\begin{equation}\label{Eq:DefHubble}
 \hubble(t,x) := \frac{e^{-\phi}}{a}\,\frac{\partial a}{\partial t} \, .
\end{equation}
Note the framework above includes all the globally hyperbolic spacetimes \cite{CMP_2005_GlobalHyperbolic}, with the Cauchy hypersurfaces $t=\textrm{const}$ identified as $\MS$, and only those, if it is compact \cite{2010_chrusciel_MathematicalGR}.

The concept of a large-scale homogeneous universe follows from the intuitive idea of tilling space in equal parts and demanding periodic boundary conditions to Einstein's equations \cite{CliftonFerreira_2009,Clifton_2012,Liu_2015, bentivegna, LGGomes_2022_CQG_2, LGGomes_2024_CQG_1} so that natural large-scale effective FLRW-like models arise after taking spatial averages \cite{LGGomes_2024_CQG_1}. Formally, we assume that all the geometric entities and their time derivatives are invariant by a free and properly discontinuous \cite{Wolf} action on $\MS$ of a discrete group $\Gamma$, so that the quotient $\K=\MS/\Gamma$ is a compact manifold. In this case, the spacetime is called a periodic cosmological model. In short, the system of PDE introduced by the Einstein's equations reduces to the spacetime $\I \times \K = \I \times (\MS/\Gamma)$, called a cosmological cell.

The Hubble evolution is isotropic for the set of observers defining the orthogonal splitting of the metric \eqref{Eq:MetricGeneral} if the trace-less part of the expansion tensor, the shear $\widehat{\sigma}$ \cite{Book_2012_ellis_mac_marteens}, vanishes. In geometric terms, that means the space sections of constant $t$ are totally umbilical hypersurfaces \cite{Book_ONeill_Semi}. This condition is equivalent to demanding the Riemaniann metric $\submetrict$ to not depend on the time $t$ (see Ref. \cite{LGGomes_2024_PeriodicHomIsotExpansion} for details),
\begin{equation}\label{Eq:AnisotropyConstraintMetric}
\widehat{\sigma} =0 
\quad \textrm{, or equivalently,} \quad
\frac{\partial{\submetrict}}{\partial t} = 0  \, .
\end{equation}
Moreover, the condition for the cosmological expansion to be homogeneous throughout space (see Ref. \cite{LGGomes_2024_PeriodicHomIsotExpansion} for details) is described by the gradient constraint involving the Hubble parameter and the Newton-like potential,
\begin{equation}\label{Eq:HomogeneityCondition}
\gradD\left( e^\phi\, \hubble \right) =0 
\quad \textrm{, or equivalently,} \quad
a=a(t) \, .
\end{equation}
It describes a universe evolving homogeneously for a set of cosmological observers, even though they perceive different rates of expansion due to their different proper-time measurements. As we add it to our assumptions, we can summarise our cosmological considerations under the following mathematical assertion:
\begin{hyp}\label{Hyp:ConstrainedEinsteinEquations}
The spacetime is a cosmological cell $\I \times \K$ with the metric 
\begin{equation}\label{Eq:MetricGeneralHomogeneousIsotropic}
 \metric = - \, e^{2 \phi(t,x)} dt^2 + a(t)^2\, \submetrict \, ,
\end{equation}
where $\gamma$ is a fixed Riemannian metric in the oriented and compact manifold $\K$ and $a:\I \times \K \to \R $ a smooth time function, $\gradD a=0$. The geometrical and physical settings are naturally brought back to $\M= \I \times \MS$ by using the covering projection $\MS \to \K = \MS/\Gamma$, in which case it turns out as a $\Gamma$-periodic cosmological model. 
\end{hyp}

We will assume the total energy-momentum content to be divided into two parts, 
\begin{equation}
T_{\mu\nu} = ^{(h)}\!\!T_{\mu\nu} + ^{(bv)}\!\!T_{\mu\nu}\, .    
\end{equation}
The first one is a perfect fluid composed of that matter-radiation counterpart which is virtually homogeneous and isotropic for the observers co-moving with the frame established by the metric (\ref{Eq:MetricGeneral}). It has the general homogeneous perfect fluid form, so that it is characterized by the "homogeneous" energy density $\energy_h$ and relativistic pressure $\pressure_h$, which are generic smooth functions of the scale factor.
As an example, in the applications to Cosmology, we might identify them with the forms usual to the $\Lambda$CDM approach \cite{Book_2012_ellis_mac_marteens}, as $\energy_h \sim \Omega_\Lambda + \Omega_r a^{-4}+ \ldots$, for instance.

The second type of component is a barotropic-viscoelastic fluid with the energy density $\energy_{(bv)}$ and relativistic pressure $\pressure_{(bv)}$ constrained by a suitable equation of state, while the anisotropic stress $\widehat{\Pi}$ satisfies the tide-curvature-stress equilibrium condition necessary for keeping the expansion isotropic along Einstein's evolution equations, that is, 
\begin{equation}\label{Eq:TideCurvatureStressEquilibrium}
\widehat{\sigma} =0 
\qquad \Rightarrow \qquad 
\widehat{\Pi} + \widehat{\Phi} = \widehat{R} \, .
\end{equation}
Here $\widehat{\Phi}$ is the traceless Newtonian-like tidal tensor (see Ref. \cite{LGGomes_2024_PeriodicHomIsotExpansion} for details) and $\widehat{R}$ the traceless part of the Ricci tensor of $a(t)^2\, \submetrict$. In the static case, $a=1$, with $\submetrict$ a metric of constant curvature, $\widehat{R}=0$, and $\phi$ weak, $|\phi|<<1$, this is the equilibrium condition $\Pi_{ik} + \widehat{\gradD_i\gradD_{k}}\phi = 0$ between the Newtonian tidal forces\cite{Book_2013_Ohanian} and the stresses in matter, where $\gradD$ is the Levi-Civita connection of $\submetrict$ and " $\widehat{\quad \cdot \quad}$ " stands for the trace-free part of any symmetric tensor of rank 2. Therefore, the relation \eqref{Eq:TideCurvatureStressEquilibrium} represents the relativistic form of the equilibrium among the internal stresses in matter, the Newtonian-like tidal forces, and the curvature of space necessary to keep the shear null. Hence, the assumption on the matter content is formulated as follows:
\begin{hyp}\label{Hyp:EnergyMomentumTensor}
Einstein's equations hold, $G_{\mu\nu}=T_{\mu\nu}$. In the referential frame of \eqref{Eq:MetricGeneral}, the net anisotropic stress satisfies the tide-curvature-stress equilibrium \eqref{Eq:TideCurvatureStressEquilibrium} while the energy density and relativistic pressure are given by $\energy (t,x) = \energy_{(h)}(a(t,x)) + \energy_{(bv)}(t,x)$ and $\pressure (t,x) = \pressure_{(h)}(a(t,x)) + \pressure_{(bv)}(t,x)$, respectively, with $\energy_{(h)}, \pressure_{(h)}: \R^*_+ \to \R$ smooth in the interval of the positive real numbers. They are constrained by an equation of state in the form
\begin{equation}\label{Eq:EquationStateGeneral}
F\left(a,x,\psi, \gradD\psi, \ldots , \gradD^k\psi \right)=0 \, ,
\end{equation}
where $k$ is a fixed integer and $F: \widetilde{W}\subset J^kE \to \R$ a smooth function defined on an open set of the $k$-jet bundle of the trivial vector bundle $E$ whose base is the spacetime $\R^*_+ \times \K$ and the fiber is the space $\R^3$, which represents the triplet $\psi=\left(\hubble,\energy_{(bv)},\pressure_{(bv)}\right)$.
%
%
\end{hyp}

The framework obtained as we gather the hypothesis H\ref{Hyp:ConstrainedEinsteinEquations} and H\ref{Hyp:EnergyMomentumTensor} is consistent with Einstein's equations. For instance, the tide-curvature-stress equilibrium condition \eqref{Eq:TideCurvatureStressEquilibrium} is exactly its trace-free spatial counterpart. Moreover, as we demand an initial condition $\hubble(t_0,x) \ne 0$ for at least one point $x \in \MS$, so that the cosmic evolution is not trivialized from the beginning, we have $\dot{a}(t_0) \ne 0$, implying that the scale factor can be used as a time coordinate. In this case, the Hubble parameter, $\hubble (a,x)$, the energy density, $\energy(a,x)$, and the relativistic pressure, $\pressure(a,x)$ should be determined by the generalized Friedmann equation \cite{LGGomes_2024_PeriodicHomIsotExpansion}, %
\begin{equation}\label{Eq:FriedmannEquation}
\frac{1}{2}(m-1)(m-2)\left( \hubble^2 + \frac{K}{a^2} \right) = 
\rho  \, ,
\end{equation}
where $(m-1)(m-2)K$ is the curvature scalar of $\submetrict$, the Raychaudhuri equation, 
\begin{equation}\label{Eq:Raychaudhuri}
\frac{\hubble}{(m-1) a^2}\, \gradD^2\left( \frac{1}{\hubble}\right) - \hubble\, a\, \frac{\partial \hubble}{\partial a} =  \frac{m-1}{2}\hubble^2 + \frac{(m-3)}{2}\frac{K}{a^2} + \frac{\pressure}{m-2} \, ,
\end{equation}
and the equation of state \eqref{Eq:EquationStateGeneral}. They form a closed system with $3$ unknowns and $3$ equations. Therefore, with $F$ suitably chosen, they should lead to well-posed problems of PDE with direct applications to Cosmology.

So far, although the scheme we have presented seems to be mathematically well established, we still have to show that it is "physically consistent", that is, we must interpret the remaining equation concerning the net energy flux of matter, represented by the vector field $q$ in $\M$. It is stated as the gradient condition (see also Ref. \cite{LGGomes_2024_PeriodicHomIsotExpansion})
\begin{equation}\label{Eq:EnergyFluxEquation}
q = \frac{(m-2)}{a^2}\,  \gradD \hubble  \, ,
\end{equation}
that is, the net energy flux follows the direction of the maximal spatial expansion. Since the law driving the behaviour of $q$ varies substantially with the subjacent model, the physical meaning of equation \eqref{Eq:EnergyFluxEquation} varies as well: convection due to the drift velocity, thermal conduction, etc. As an example, if we assume that the barotropic-viscoelastic fluid is responsible for the net energy flux through heat conduction, we should look at the equation \eqref{Eq:EnergyFluxEquation} as connecting the thermodynamic variables (local temperature, entropy density, chemical potential, etc) with $\hubble$. There are many possibilities which are physically consistent with our framework \cite{Book_2012_ellis_mac_marteens, LRGR_2007_RelativisticFluidMechanics}, and the reader may check that some of them can be easily handled\footnote{As an example, we could pick the typical non-causal law for the heat flow as stated in sec. 5.2 of Ref.\cite{Book_2012_ellis_mac_marteens}. This would give an explicit form for the temperature as a function of $\hubble$ and the entropy flow density vector would satisfy the second law $\grad_\mu S^\mu \ge 0$ under usual conditions for the bulk pressure during expansion. }. Here we refrain from developing them any further and content ourselves by claiming that the study of any closed system in the form of the equations \eqref{Eq:EquationStateGeneral},  \eqref{Eq:FriedmannEquation}, and \eqref{Eq:Raychaudhuri} can be straightforwardly converted to interesting and physically relevant application to Cosmology as far as $F$ is meaningful for that context. Therefore, we can finally state our general mathematical problem as follows:
\begin{itemize}
\item {\bf The H1+H2 Evolution Problem:}
For a given compact Riemannian manifold $\K$ with metric $\submetrict$, investigate the $a$-time evolution of the functions $\hubble$, $\energy$ and $\pressure$ on $\K$ driven by the equations \eqref{Eq:EquationStateGeneral},  \eqref{Eq:FriedmannEquation}, and \eqref{Eq:Raychaudhuri}. The initial condition for $\hubble$ at $a=a_0$ must be everywhere positive (or negative).    
\end{itemize}
%

At this point, it is important to set the right spot to where our H1+H2 Evolution Problem belongs. The reader used to the Einstein's equations in the Mathematical General Relativity literature \cite{Book_choquet_Bruhat, 2010_chrusciel_MathematicalGR, 2005_LRR_Existence_Theorems_GR, 2017_Coley_OpenProblemsGR, Christodoulou_MinkowskiStability_1993} may find our proposal and notation quite unfamiliar, so some further explanation seems necessary. First of all, our system is severely constrained, as the title suggests. Besides the Einstein constraint equations \eqref{Eq:FriedmannEquation} and \eqref{Eq:EnergyFluxEquation}, the so-called tide-stress-curvature equilibrium \eqref{Eq:TideCurvatureStressEquilibrium} is much more restrictive, such that we should not even expect to obtain hyperbolic evolution PDEs from our hypotheses. However, assertions like that are necessary to balance the extent to which we might probe the mathematical aspects of General Relativity without losing the working cosmologist from our sight. In this perspective, hypotheses H\ref{Hyp:ConstrainedEinsteinEquations} and H\ref{Hyp:EnergyMomentumTensor} are justifiable in many ways. Here we put down some advantages of investigating the periodic cosmological models arising from them: (i) They  contain any FLRW or any spatially periodic generalized RW spacetime and provide a quite simple way to generate examples of interest in both Mathematics and Cosmology; (ii) Understanding the evolution of the spacetimes satisfying Hypothesis \ref{Hyp:ConstrainedEinsteinEquations} is a central problem in the mathematical foundations of Cosmology, since it is closely related to its main tenet, the cosmological principle; (iii) They represent the asymptotic future behaviour of the spatially periodic class within many spacetimes of cosmological interest \cite{1983_PRD_Wald1_CosmicNoHair, CMP_2020_Creminelli, 2024_AHP_Friedrich, 2024_PTRSA_Friedrich_Asymptotic_Cosmology}; (iv) They give a relatively simple framework suitable to generate examples where we can understand the non-linear behaviour of the inhomogeneities along the cosmic evolution; (v) There is a natural way to build effective $\Lambda$CDM-like models from them, just as suggested in Ref. \cite{LGGomes_2024_CQG_1}, by taking large-scale averages of any $\Gamma$-periodic function $f:\M \to \R$, which can be readily identified as a function defined on $\K$, through
\begin{equation}\label{Eq:AverageValue}
\langle f \rangle (t):= \frac{1}{L_0^{m-1}}\int_\K f(t,x) \sqrt{\det \submetrict}\, d^{m-1}x\, ,
\end{equation}
where $\sqrt{\det \submetrict}\, d^{m-1}x$ is the natural $\submetrict$-volume form of $\K$ and $L_0^{m-1}$ the m-volume of $\K$. So, they can be readily used by the working cosmologist.

In the rest of the manuscript, we will investigate the H1+H2 Evolution Problem with the equation of state \eqref{Eq:EquationStateGeneral} given by 
\begin{equation}
\pressure_{(bv)} =  \eqstate(a) \energy_{(bv)} - \, (m-1)\,  \zeta(a) \,\hubble^2 
\, ,
\end{equation}
where $\eqstate$ is the barotropic and $\zeta$ the dimensionless bulk viscosity coefficients \footnote{We have opted to introduce only the most natural and physically relevant dimensionless parameters, as $\zeta\, \hubble$ instead of $\zeta$, for instance (Compare with section 5.2 of Ref. \cite{Book_2012_ellis_mac_marteens}). Of course, many other possibilities would not substantially change our overall approach. See Ref. \cite{CQG_1995_Maartens_DissipativeCosmology} for many examples of different equations of state.}.
Formally, we investigate the equations \eqref{Eq:FriedmannEquation} and \eqref{Eq:Raychaudhuri} subjected to the constraint
\begin{equation}\label{Eq:EquationStateViscousFluid}
\pressure = \pressure_{h}(a) + \eqstate(a) \left(\energy - \energy_h (a) \right) -\,(m-1)\, \zeta(a)\, \hubble^2 
\, .
\end{equation}
where $\energy_h, \pressure_{h}, \eqstate,  \zeta: \R^*_+ \to \R$ are smooth functions. We set the initial value problem (IVP) related to it in section \ref{Sec:IVP}, for which we prove existence, uniqueness, and stability in section \ref{Sec:MainTheorem}. In sections \ref{Sec:PeriodicCosmologicalModels} and \ref{Sec:InequalitiesEnergyDensity}, we return to the associated periodic cosmological model by translating the results of section \ref{Sec:MainTheorem} to this context. In particular, we obtain general results concerning not only the existence and uniqueness of the solutions,  but even further, we also set bounds for the evolution of inhomogeneities and show a new mechanism of energy transfer occurring between the viscoelastic fluid and the homogeneous expansion. In section \ref{Sec:EffectiveStandardModel}, we apply those findings to Cosmology under the light of the  $\Lambda$CDM-like approach. In particular, a counter-intuitive new term appears in the averaged energy density, namely,    
\begin{equation}\label{Eq:EnergyAveragedGeneral}
\langle \energy\rangle = \underbrace{\left(\Omega_\Lambda + \frac{\Omega_m}{a^3} +\frac{\Omega_r}{a^4}\right)\energy_c}_{\Lambda \textrm{CDM}}  
\quad + \quad  
\underbrace{\energy^{\textrm{eff}}(a)}_{\textrm{New}} 
\quad + \quad   \underbrace{\frac{\langle \delta\energytt\rangle }{a^{3}}}_{\textrm{"small"}} \, ,
\end{equation}
which is a consequence of corollary \ref{Thm:EnergyDensity}  (see sections 6 and 7 of Ref. \cite{LGGomes_2024_CQG_1}) and is of great interest in observational Cosmology. Finally, by the end of section \ref{Sec:EffectiveStandardModel}, we find some exact solutions of the IVP and use them to exemplify the general outcomes obtained previously and their relations to the periodic cosmological models arising in such a context. We finish our manuscript with some final remarks in section \ref{sec:Final Remarks}. Throughout the manuscript, we use Einstein's convention for sums, with Greek indexes running as $\mu,\nu, \kappa =0, \ldots, m-1$ and Roman ones as $i,j,k=1, \ldots, m-1$. The conventions on signs are the same as in Ref. \cite{Book_2012_ellis_mac_marteens}.

\section{The initial value problem arising from the  H1+H2 Evolution: an example}
\label{Sec:IVP}

Led by the generalized Friedmann equation \eqref{Eq:FriedmannEquation}, Raychaudhuri's equation \eqref{Eq:Raychaudhuri}, and the equation of state \eqref{Eq:EquationStateViscousFluid}, we adopt some new parameters as
:
\begin{equation}\label{Eq:ParameterXi}
\xi(a)=\frac{m-1}{2} \int_{a_0}^{a}\left(1+\eqstate(b)- \frac{2 \zeta(b)}{m-2}\right)
\frac{db}{b} \, ,
\end{equation}
%
%
%
%
\begin{eqnarray} \label{Eq:BetaFunction}
\beta(a,x) &=& 
- \, \frac{m-1}{m-2}\left(\pressure_h-\eqstate \energy_h\right)  a^2 \nonumber  - \frac{m-1}{2}\left( m-3 + (m-1)\eqstate \right)K\, ,
\end{eqnarray}
and 
\begin{equation}\label{Eq:DefinitionHubblet}
\hubblet = \frac{1}{e^{\xi(a)}\hubble}\, ,
\end{equation}
together with the new time coordinate
\begin{equation}\label{Eq:DefinitionTimeParameter}
s(a) =  -\, \frac{1}{m-1}\int_{a_0}^{a}\, e^{2 \xi(b)}\frac{db}{b^3} 
\, .
\end{equation}
%
%
%
Under this new set of parameters, we are led to the initial value problem (IVP):
\begin{itemize}
    \item {\bf The IVP:} \emph{ Given a compact and connected Riemannian Manifold $(\K,\submetrict)$ with the Laplace-Beltrami operator $\gradD^2$ and a smooth function $\beta: \mathsf{J} \times \K \to \R$, $\mathsf{J} \subset \R$ an open interval containing $0$, we look for a continuous function $\hubblet: [0, s_{max}) \times \K \to \R$, smooth in $(0, s_{max}) \times \K$, satisfying the PDE
    \begin{equation}\label{Eq:RestrictedEinsteinFormal}
    \frac{\partial }{\partial s}\hubblet  =  \hubblet^2 \gradD^2 \hubblet + \beta(s,x) \hubblet^3 , \; \; \mbox{on} \; (0, s_{max}) \times \K \, ,
    \end{equation}
    with the initial condition
    \begin{equation}\label{Eq:RestrictedEinsteinFormalIC}
    \forall\, x \in \K \quad : \quad \hubblet(0,x)  =  \hubblet_0(x) \ne 0 \, ,
    \end{equation}
    for a given smooth function $\hubblet_0: \K \to \R$ vanishing nowhere in $\K$.}
\end{itemize}

Any solution of this IVP leads us to a periodic cosmological model with homogeneous and isotropic Hubble evolution according to the arguments put forth in the introduction. Hence, in the rest of the manuscript, we shall exploit this IVP and connect our results with their cosmological implications.

\section{The main theorems}
\label{Sec:MainTheorem}
%

We begin by proving existence, uniqueness, and continuous dependence on the initial conditions for the IVP defined at the end of section \ref{Sec:IVP}: 
\begin{thm}\label{Thm:ExistenceUniqueness} 
%
There is a value $0< s_{max} \le \infty$, $[0,s_{max}) \subset \mathsf{J}$, and a unique continuous function $\hubblet :[0,s_{max}) \times \K \to \R$, smooth in $(0,s_{max}) \times \K$, satisfying the initial value problem IVP defined at the end of section \ref{Sec:IVP}. We assume $s_{max}$ to be maximal with these properties. The mapping $ \hubblet_0   \mapsto \hubblet $ is locally continuous \footnote{Here, we assume the usual Fr\'echet space topology of smooth functions on compact manifolds \cite{Book_2011_Saunders} for which the subset $C^\infty_{\ne 0}(\K)$ of the initial data satisfying (\ref{Eq:RestrictedEinsteinFormalIC}) is open in $C^\infty(\K)$. Formally, for any $\hubblet_0 \in C^\infty_{\ne 0}(\K)$ and $0\le T < s_{max}$ there is an open set $U_0 \in \mathcal{V}_{T} \subset C^{\infty}_{\ne 0}(\K)$ such that any solution with the initial condition in $\mathcal{V}_T$ is defined in $[0,T]\times \K$ and the mapping $\hubblet_0  \in \mathcal{V}_T  \mapsto \hubblet \in C^{\infty}([0,T] \times \K)$ is continuous.}. Furthermore, as we fix $0 \le T < s_{max}$ and define
\begin{equation}\label{Eq:ConstantsMaxMin}
 \omega = \min_{x\in \K}|\hubblet_0(x)| \qquad \textrm{and} \qquad C(T) = \omega^2 \min_{[0,T]\times\K}\beta(s,x) \, ,
\end{equation}
the solution satisfies 
\begin{equation}\label{Eq:InequalityThmExistenceUniqueness}
\omega \, e^{C(T)\, s}  \le |\hubblet(s,x)|  \, , 
\end{equation}
for all $0\le s\le T$ and $x \in \K$.
\end{thm}
\proof
Before assuring existence, let us prove that inequality (\ref{Eq:InequalityThmExistenceUniqueness}) holds for any solution of the aforementioned IVP. For this, assume that $\hubblet:[0,s_{max})\times \K \to \R$, $s_{max}$ is one of them represented in its maximal domain of definition. Since $\hubblet_0\ne 0$ in $\K$, the PDE (\ref{Eq:RestrictedEinsteinFormal}) is invariant under the transformation $\hubblet \mapsto - \hubblet$, and $\K$ is connected, we can assume $\hubblet_0(x)>0$, with no loss of generality. Fix $0 < T < s_{max}$, the constants $\omega > 0$ and $C=C(T)$ as in the equation (\ref{Eq:ConstantsMaxMin}), and $C^*<C$. We can pick $0<\epsilon_1<\omega$ such that $C^* < \beta(s,x)(\omega -\epsilon)^2$ for all $0 \le \epsilon < \epsilon_1$, $0 \le s \le T$, and $x \in \K$, that is,  
\begin{equation}\label{Eq:ProofTheoremConstantC*}
e^{-2C^*\, s}C^* \le C^*  < \beta(s,x)(\omega -\epsilon)^2\, .
\end{equation}
 With $s$ and $\epsilon$ in those intervals, define the function 
\begin{equation}
F_\epsilon(s,x)= \hubblet(s,x) + (\epsilon - \omega)e^{C^*\, s} \, .
\end{equation}
Suppose it has the first zero at $(s_1,x_1)$, $s_1 < T$, that is, $F_\epsilon(s,x)$ has no zeros for $0\le s < s_1$. Since $F_\epsilon(0,x) = \hubblet_0(x)-\omega + \epsilon >0$, we must have $\partial F_\epsilon/\partial s \le 0$ in $(s_1,x_1)$, that is, $F_\epsilon$ must be decreasing in $s$ at this point. Furthermore, $\hubblet(s_1,x_1)=(\omega-\epsilon)e^{C^*\,s_1}$. So, in the point $(s_1,x_1)$ we obtain   
\begin{eqnarray*}
   \frac{\partial F_\epsilon}{\partial s} &=& \frac{\partial \hubblet}{\partial s}  +  C^* (\epsilon - \omega) e^{C^*\, s_1} \\
&=& \hubblet^2 \gradD^2 \hubblet +  \beta(s_1,x_1) \hubblet^{3} +  C^* (\epsilon - \omega) e^{C^* s_1}. \\
&=& (\omega-\epsilon)^2 e^{2 C^* s_1} \gradD^2\hubblet +  (\omega-\epsilon) e^{3 C^* s_1}\left( \beta(s_1,x_1) (\omega-\epsilon)^{2} -  C^* e^{-2 C^* s_1}\right)  \le   0 . 
\end{eqnarray*}
According to the definition of $C^*$ in eq. (\ref{Eq:ProofTheoremConstantC*}), this inequality is possible only if $\gradD^2\hubblet(s_1,x_1)=\gradD^2 F_\epsilon(s_1,x_1) <0$, which is a contradiction, for $x_1$ is a minimum in $\K$ of the map  $x \mapsto F_\epsilon(s_1,x)$, and thus we must have $\gradD^2 F_\epsilon(s_1,x_1) \ge 0$. Therefore, there is no point $(s,x)$ with $0\le s<T$ where $F_\epsilon(s,x)=0$ for any $0 \le \epsilon < \epsilon_1$, that is, $\hubblet(s,x) \ge \omega e^{C^*\, s}$ for any $0 \le s \le T$ and $C^* < C$. Therefore, $\hubblet(s,x) \ge \omega e^{C\, s}$  in $[0,T]\times \K$, proving (\ref{Eq:InequalityThmExistenceUniqueness}).

Returning to the existence and uniqueness, we note that in Ref. \cite{MM2012}, the authors have investigated the IVP on $\K$ driven by the PDE
\begin{equation}\label{Eq:EDPGeneralParabolic}
\frac{\partial \hubblet}{\partial s} = Q^i_k(t, x, \hubblet , \gradD \hubblet) \gradD_{i}\gradD^k\hubblet + b(t, x, \hubblet, \gradD \hubblet) \, ,
\end{equation}
where all the functions involved are smooth in their arguments \footnote{Formally, we consider the trivial line bundle  $\R \times \K$, for which every real-valued function on $\K$ is identified as a section, and its first jet bundle $J(\R \times \K)$. Hence, $b$ is a smooth real-valued function on $J(\R \times \K)$ and $Q:J(\R \times \K) \to \mathcal{T}^1_1(\K)$ a $\K$-bundle morphism on the bundle of the $(1,1)$-tensors.} with $Q^i_k$ positive definite with the lowest eigenvalue uniformly bounded away from zero. In this case, the IVP exists for $m \geq 3$, is unique, and depends continuously on initial data in the $C^\infty$-topology (see Appendix A in Ref. \cite{M2010}). For $m=2$, we have $\K=S^1$, and the existence and uniqueness is assured by the results of Ref. \cite{TaylorIII} (see chapter 15).

Unfortunately, we cannot apply this argument straightforwardly to our problem, for the eigenvalues of $Q^k_i= \hubblet^2 \delta^k_i$ are not bounded away from zero. Hence, keeping the constants fixed as before, define 
\begin{equation}
\widetilde{\omega} = 
\renewcommand\arraystretch{1.5} 
\left\{
\begin{array}{ll}
 \omega & , \; \mbox{if}\;  C \geq 0\\
 \omega\, e^{C T}  & , \;  \mbox{if} \;  C < 0  \\
\end{array}
\right. 
\qquad , \qquad  
m_{\omega} (\hubblet) = 
\renewcommand\arraystretch{1.5} 
\left\{
\begin{array}{ll}
 \hubblet^2 & , \; \mbox{if}\;  \hubblet \geq \widetilde{\omega} /2\\
 \widetilde{m}_{\omega} (\hubblet)  & , \;  \mbox{if} \;  \hubblet < \widetilde{\omega}/2 \\
\end{array}
\right. \, ,
\end{equation}
and $Q_i^k = m_{\omega}(U) \delta_i^k$ , 
%
%
where $\widetilde{m}$ is chosen such that $m_\omega(\hubblet)$ is smooth in both entries and $\widetilde{m}_{\omega} (\hubblet) \geq \widetilde{\omega}^2/8$. Since $\omega$ and $\widetilde{\omega}$ are positive, the operator $Q_{ij}$ defined above satisfies, for every spatial vector $\xi$,  
\[
Q_{ij}\xi^{i} \xi^{j} = m_{\omega }(U) \xi^2 \geq \frac{\widetilde{\omega}^{2}}{8} \xi^2 \, . 
\]
Therefore, it is strongly elliptic with eigenvalues uniformly bounded away from zero, allowing the existence of a unique solution $\hubblet^{(T)}$ defined in the manifold $[0, s_{max}^{(T)}) \times \K$ depending continuously on the initial condition $\hubblet_0$ in the $C^{\infty}$-topology, where $s_{max}^{(T)}$ is taken to be maximal with the constraint $s_{max}^{(T)} \le T$. Since $\K$ is compact, at least in a small interval $[0, s^*)$, $s^* \le s_{max}^{(T)}$ by continuity and the definition of $\omega$, we have $\hubblet^{(T)} \ge \widetilde{\omega}/2$, so that $m_{\omega} (\hubblet^{(T)})=(\hubblet^{(T)})^2$. In other words, $\hubblet^{(T)}$ is our desired solution $\hubblet$, at least in $[0, s^*)$. As before, let $[0,s_{max})\times \K$ denote the maximal domain for $\hubblet$.  

Suppose that $s^* < s_{max}^{(T)} \le T$, with $s^*$ maximal such that $\hubblet^{(T)} = \hubblet$ in $[0, s^*)\times \K$. By the property (\ref{Eq:InequalityThmExistenceUniqueness}), for any $0< T' <s^*$, we have $\omega \, e^{C(T')\, s}  \le \hubblet^{(T)}(s,x)$ for $0 \le s \le T'$. Since $T' < T$, $C(T) \le C(T')$, we also have $\omega \, e^{C(T)\, s}  \le\omega \, e^{C(T')\, s}  \le \hubblet^{(T)}(s,x)$. If $C(T)\ge 0$, we have $\hubblet^{(T)}(s,x) \ge \omega = \widetilde{\omega}$. If $C(T)< 0$, we have $\hubblet^{(T)}(s,x) \ge \omega e^{C(T)T} = \widetilde{\omega}$. In resume,  $\hubblet^{(T)}(s,x) \ge \widetilde{\omega}$ for any $0 \le s \le s^*$. But this implies that $m_{\omega} (\hubblet^{(T)}(s,x))=\left(\hubblet^{(T)}(s,x)\right)^2$ in $[0,s^{**})\times \K$ for some $s^{**} > s^{*}$, thus extending the equality  $\hubblet^{(T)} = \hubblet$ to the same set. Since this contradicts our assumption, we must have $s^*=s_{max}^{(T)}$, that is, $s_{max}^{(T)} \le s_{max}$.  

We conclude that, if $T < s_{max}$, then $s^*=s_{max}^{(T)}=T$. Therefore, uniqueness of $\hubblet^{(T)}$ in $[0, T)\times \K$ is extended to $\hubblet$ in the same domain. Since it holds for any $T < s_{max}$, $\hubblet$ is unique. Moreover, the result in Appendix A of Ref. \cite{M2010} guarantees not only existence and uniqueness but also continuous dependence on initial data, so that $U_0 \mapsto \hubblet^{(T)}=\hubblet$ is locally continuous in the sense explained before, thus completing the proof of our theorem.
\qed

Before returning to the physical variables, we shall comment on possible extensions to theorem \ref{Thm:ExistenceUniqueness}. In Ref. \cite{TaylorIII}, many issues related to solutions to the Cauchy problem for the quasi-linear PDE are addressed. For instance, in Chapter 15, Sec. 7, the IVP on the torus $\K = \mathbb{T}^{m-1}$ given by $u(0, x) = u_0(x)$ and
\begin{equation}\label{Eq1}
 \partial_{s} u = A^{jk}(t, x,u, \partial_{x}u) \partial_{j} \partial_{k} u + B(t, x,u, \partial_{x}u)   \; \; , 
\end{equation}
with $A^{jk}$ and $B$ smooth in their arguments, the first one satisfying a strong parabolicity condition, allows the local existence of solutions for $u_0 \in H^{\ell}(\K)$, with $\ell > \frac{m-1}{2} + 1$. Using a modified Galerkin method, it is possible to prove that a solution $u$ exists on an interval $I = [0, T)$ such that $u \in L^{\infty}(I, H^{\ell}(\K)) \cap Lip(I, H^{\ell - 2}(\K))$. Moreover, the proposition 7.4 (p. 382) in Ref. \cite{TaylorIII} guarantees that  $u \in C^{\infty}((0, T) \times \K)$. Indeed, the regularity of the initial data in Sobolev spaces $H^{\lambda}(\K)$  can be interpreted in terms of the $C^{r}(\K)$ regularity as follows: in a compact Riemannian $(m-1)$-manifold $(\K,\submetrict)$, $q \geq 1$ and $r < \lambda$ two integers, if $ \lambda > r + \frac{m-1}{q}$ then $W^{q, \lambda}(\K) \subset C^{r}(\K)$ (see Ref. \cite{hebey2000nonlinear}, p.37). In particular, if $\lambda > r + \frac{m-1}{2}$, we have $H^{\lambda}(\K) \subset C^{r}(\K)$.

In Section 8 of Ref. \cite{TaylorIII}, sharper results in the case of general compact manifolds are obtained from the para-differential operator calculus. With $A^{jk}=A^{jk}(t, x, u)$ and $B=B(t, x,u)$, for $u_0 \in H^{\ell}(\K)$,  $\ell > \frac{m-1}{2} + 1$, there is a unique solution satisfying $u \in C([0, T], H^{\ell}(M)) \cap C^{\infty}((0, T) \times M)$,  which persists as long as $\|u(t)\|_{\mathcal{HC}^{r}}$ is bounded, given $r >0$, where $\mathcal{HC}^{r}$ denotes the space of Hölder-continuous functions on $M$.

Summing up the comments above, we could have approached theorem \ref{Thm:ExistenceUniqueness} with the initial conditions in a broader context, which would soon regularize to a smooth function on $\K$ for $s>0$. However, we have opted to work in the $C^\infty$ perspective because it simplifies our assumptions and is good enough for cosmological applications.

The inequalities obtained in theorem \ref{Thm:ExistenceUniqueness} can be improved as follows:
\begin{thm}\label{Thm:InequalitiesMaxMin} 
Any solution $\hubblet :[0,s_{max}) \times \K \to \R$ of the aforementioned IVP satisfies the inequalities   
\begin{equation}\label{Eq:InequalityMaxMin}
\frac{1}{\Omega^2}
\le \frac{1}{\hubblet(s,x)^2} + 2  \int_0^s\beta(u,x)du \le
\frac{1}{\omega^2} \, ,
\end{equation}
for any $0 \le s < s_{max}$ and $x \in \K$, where 
\begin{equation}\label{Eq:ConstantsMaxMin2}
 \Omega = \max_{x\in \K}|\hubblet_0(x)| \qquad \textrm{and} \qquad \omega = \min_{x\in \K}|\hubblet_0(x)|\, .
\end{equation}
\end{thm}
\proof 
As in theorem \ref{Thm:ExistenceUniqueness}, assume $\hubblet_0>0$ with no loss of generality. By the property (\ref{Eq:InequalityThmExistenceUniqueness}), $\hubblet >0$. Let $\omega$ and $\Omega$ be the positive constants introduced in eq. (\ref{Eq:ConstantsMaxMin2}) and define $0<s^* \le s_{max}$ such that  
\begin{equation}
2  \int^{s}_0\, \beta(u,x)du < \frac{1}{\Omega^2}  \quad , \quad 0\le s < s^*\, , \, x \in \K \, ,
\end{equation}
which exists by the compactness of $\K$.
For each $0<\tau<s^*$, we can find $\epsilon_1>0$ for which  
\begin{equation}
\chi(s,x) = \left( 1 - 2 \epsilon s - 2 (\Omega+\epsilon)^2 \int_0^s\beta(u,x)du \right)^{-1/2}
\end{equation}
is well defined and smooth for every $0 \le \epsilon < \epsilon_1$, $0 \le s \le \tau$, and $x \in \K$. Moreover, the function
\begin{equation}
G_\epsilon(s,x)=  (\Omega+\epsilon)\chi(s,x) - \hubblet(s,x) 
\end{equation}
satisfies $G_\epsilon(0,x) = \Omega + \epsilon-\hubblet_0(x) > 0$. Assuming $(s_1,x_1)$ is its first zero, $s_1 < \tau$, that is, $G_\epsilon(s,x)$ has no zeros for $0\le s < s_1$, then the conditions $\hubblet(s_1,x_1)=(\Omega+\epsilon)\chi(s_1,x_1)$ and $\partial G_\epsilon/\partial s (s_1,x_1)\le 0$ must hold. Analyzing them in the point $(s_1,x_1)$, we obtain   
\begin{eqnarray*}
\frac{\partial G_\epsilon}{\partial s} &=& (\Omega+\epsilon)\frac{\partial \chi}{\partial s} - \frac{\partial \hubblet}{\partial s}  \\
&=& \left(\epsilon (\Omega+\epsilon) + (\Omega+\epsilon)^3\beta(s_1,x_1)\right) \,\chi(s_1,x_1)^3 
-  \beta(s_1,x_1) \hubblet^{3} - \hubblet^2 \gradD^2 \hubblet  \\
&=& \epsilon (\Omega+\epsilon)\,\chi(s_1,x_1)^3 + (\Omega+\epsilon)^2\chi(s_1)^2 \gradD^2G_\epsilon (s_1,x_1) \\
&\le & 0 \, .  
\end{eqnarray*}
This implies the contradiction $\gradD^2 G_\epsilon(s_1,x_1) <0$, for $x_1$ is a minimum in $\K$ of the map  $x \mapsto G_\epsilon(s_1,x)$, and thus we must have $\gradD^2 G_\epsilon(s_1,x_1) \ge 0$. Therefore, there is no point $(s,x)$ where $G_\epsilon(s,x)=0$ for any $0 \le \epsilon < \epsilon_1$, that is, for any $0 \le s \le s^*$ and $x \in \K$, 
\begin{equation}
\hubblet(s,x) \le \frac{\Omega}{\sqrt{1 - 2 \Omega^2 \int_0^s\beta(u,x)du}} \, ,
\end{equation}
that is,
\begin{equation}
\frac{1}{\Omega^2} - 2  \int_0^s\beta(u,x)du
\le \frac{1}{\hubblet(s,x)^2}\, . 
\end{equation}
This inequality holds even when the left-hand side is negative so that we can put $s^*=s_{max}$. 

As we repeat the same argument with the following modifications, 
\[
\Omega \to \omega \quad , \quad 
\chi(\epsilon,s) \to \chi(-\epsilon,s) 
\quad , \textrm{and} \quad 
G_\epsilon(s,x) \to - G_{-\epsilon}(s,x) \, ,
\]
with $s^{**}$ maximal such that
\begin{equation}
2  \int^{s}_0\, \beta(u,x)du < \frac{1}{\omega^2}  \quad , \quad 0\le s < s^{**}\, , \, x \in \K \, ,
\end{equation}
there we must have 
%
\begin{equation}
\frac{\omega}{\sqrt{1 - 2 \omega^2 \int_0^s\beta(u,x)du}} \le \hubblet(s,x) \, . 
\end{equation}
It also shows that $s^{**}=s_{max}$, otherwise the solution would blowup at $s=s^{**}$. Therefore,
\begin{equation}
\frac{1}{\hubblet(s,x)^2} \le
\frac{1}{\omega^2} - 2  \int_0^s\beta(u,x)du  
\end{equation}
for $0 \le s < s_{max} \, , \, x \in \K$, thus proving the theorem.
\qed
%
%

\section{Existence, uniqueness, and relative stability for the 
periodic cosmological models}
\label{Sec:PeriodicCosmologicalModels}

Now we are in the position to prove that our whole scheme presented in the introduction with the physical fluid put forth in section \ref{Sec:IVP} is well-posed. We start by formalizing our given data:
\begin{itemize}
    \item {\bf Structural data:} $\MS$ is a connected manifold and $\Gamma$ is a discrete group acting freely and properly discontinuously on it. The quotient manifold $\K = \MS/\Gamma$ is compact and oriented. Any tensor field in $\K$ can be raised to one of the same kind in $\MS$ through the pull-back with the canonical projection $S \to \K$, since it is a local diffeomorphism. They are called $\Gamma$-periodic tensor fields in $\MS$. The parameters $\energy_h$, $\pressure_h$, $\eqstate$, and $\zeta$ are all given real-valued time functions defined in an open interval $\mathsf{J} \subset \R$ containing $0$, so that the smooth function $\beta: \mathsf{J}\times \K \to \R$ defined by the relation (\ref{Eq:BetaFunction}) is $\Gamma$-periodic whenever the Riemannian metric $\submetrict$ is as well.       
\end{itemize}
The theorem of existence and uniqueness for solutions follows as:
\begin{cor}\label{Thm:ExistenceUniquenessHubble} 
Under the hypotheses of the aforementioned structural data, let $\gamma$ be a Riemannian metric in $\MS$ and $\hubble_0:\MS \to \R$ a smooth and nowhere vanishing function, both $\Gamma$-periodic. Given $a_0 > 0$, there are a minimal value $a_{min}$ with $0 \le a_{min} < a_0$ and continuous $\Gamma$-periodic functions $\hubble, \energy, \pressure:(a_{min},a_0] \times \MS \to \R$, smooth in $(a_{min},a_0) \times \K$, which are uniquely characterized by satisfying Friedmann's (\ref{Eq:FriedmannEquation}) and Raychaudhuri's (\ref{Eq:Raychaudhuri}) equations, the constraint (\ref{Eq:EquationStateViscousFluid}) imposed by the equation of state of the fluid of section \ref{Sec:IVP}, and the smooth initial condition $\hubble(a_0,x)=\hubble_0(x)\ne 0$, $x \in \K$. As we fix $a_{min}<a_1\le a_0$ and define the constants 
\begin{equation}\label{Eq:ConstantsMaxMinHubble}
\hubble_M = \max_{x\in \K}|\hubble_0(x)| 
\qquad \textrm{and} \qquad 
C(a_1) = \frac{1}{\hubble_M^2} \min_{[a_1,a_0]\times\K}\beta(a,x) \, ,
\end{equation}
we have for any $a_1<a\le a_0$ and $x \in \K$
\begin{equation}
|\hubble(a,x)| \le \hubble_M \, e^{-\left(\, \xi(a)+ C(a_1)\, s(a)\, \right)} \, .   
\end{equation}
The spacetime $\M=(a_{min},a_0) \times \K$ with the metric 
\begin{equation}\label{Eq:MetricHubbleForm}
 \metric = - \, \frac{da^2}{a^2\, \hubble^2} + a^2\, \submetrict 
\end{equation}
is a periodic cosmological model with the discrete group of symmetries $\Gamma$, which is uniquely defined up to an isometry. Moreover, the solutions depend continuously on the initial data in the $C^{\infty}$-topology for the $\Gamma$-periodic functions, which we identify with functions on the compact manifold $\K=\MS/\Gamma$.

\end{cor}
\proof
The corollary follows from the identity (\ref{Eq:DefinitionHubblet}) and theorem \ref{Thm:ExistenceUniqueness}, since $\hubble_0\ne 0$ in $\K$ means that $\hubblet_0$ has no zeros, the initial condition with $a=a_0$, that is, $s=0$. 
\qed

Corollary \ref{Thm:ExistenceUniquenessHubble} generalizes theorem 1 in \cite{LGGomes_2022_CQG_2} in two important points: the spatial sections have any geometry whatsoever, rather than being flat, and the class of the allowed fluids is much larger, as we have seen in section \ref{Sec:IVP}. On the other hand, we have adopted the stronger assumption of smoothness, contrary to the $C^k$-differentiability of the former, which has already been discussed just after the proof of the theorem \ref{Thm:ExistenceUniqueness}.

\section{Inequalities for the energy density}           %
\label{Sec:InequalitiesEnergyDensity}                                                 %

\subsection{The decomposition of the energy density}
\label{Sec:DecompositionEnergyDensity}

Let us now investigate the behaviour of the energy density in our periodic models. Our main result is:     
\begin{cor}\label{Thm:EnergyDensity}
Let be given a solution of Einstein's equations as in corollary \ref{Thm:ExistenceUniquenessHubble} and write 
\begin{equation}\label{Eq:EnergyDensityResidual}
\energy(a,x) = \energy_h(a)+ \energy_v(a)+ \energyttt(a) + \energytt (a,x) \, e^{-2 \xi(a)} 
\, ,
\end{equation}
where
\begin{equation}\label{Eq:EnergyDensityViscosity}
\energy_v =   \frac{2(m-1)}{m-2}\, e^{-2 \xi(a)} \, \int_{a_0}^a\, e^{2 \xi(u)}\zeta(u)\, \energy_h(u)\frac{du}{u} 
\end{equation}
is the viscosity energy density,
\begin{equation}\label{Eq:EnergyDensityTransfer}
\energyttt(a) =   -\, e^{-2 \xi(a)} \, \int_{a_0}^a\, e^{2 \xi(u)}\energyt (u)\, du \, ,
\end{equation}
with $\energyt(a)$ the transfer-energy density defined as
\begin{equation}\label{Eq:EnergyDensityResidualInequalitiesQ}
\frac{d\energy_h}{da} + \frac{(m-1)}{a}\left(\energy_h+\pressure_h \right)= \energyt \, , 
\end{equation}
and $\energytt (a,x)$ is the bulk energy density uniquely defined by the relations above. It is bounded by
\begin{equation}\label{Eq:EnergyDensityResidualInequalities}
\mathsf{C}_m\left(\hubble_{min}^2 - \hubble_0^2\right) + \widetilde{K}
\le \energytt - \energytt_0 \le 
\mathsf{C}_m\left(\hubble_{max}^2 - \hubble_0^2\right) + \widetilde{K}
 \, ,
\end{equation}
whenever $a_{min}<a\le a_0$ and $x \in \MS$, with $\energytt_0(x)=\energytt(a_0,x)$, $\mathsf{C}_m =\frac{(m-1)(m-2)}{2}$, and 
\begin{equation}
\widetilde{K}(a,x):=-\, (m-1)^2 \, K(x)\,  \int_{a_0}^a\, e^{2 \xi(u)}\frac{\zeta(u)}{u^3}\, du 
 \, ,
\end{equation}
where
\begin{equation}\label{Eq:ConstantsMaxMin3}
 \hubble_{max} = \max_{x\in \K}|\hubble_0(x)| \qquad \textrm{and} \qquad \hubble_{min} = \min_{x\in \K}|\hubble_0(x)|\, .
\end{equation}
\end{cor}
\proof
Firstly, we note that, by using the definitions above, 
\begin{equation*}
\beta ds  
= -\, \frac{\beta}{m-1} \frac{e^{2 \xi}}{a^3}da
= ds_1 + \frac{K}{2}d\left( \frac{e^{2 \xi}}{a^2} \right) + \zeta\, \frac{(m-1)Ke^{2 \xi}}{(m-2)a^3}da  \, ,    
\end{equation*}
where
\begin{eqnarray*}
s_1(a) 
&=& \int_{a_0}^{a}\frac{\pressure_h-\eqstate \energy_h}{m-2}\frac{e^{2 \xi(u)}}{u}du \\
&=& \frac{1}{2\mathsf{C}_m} \, \int_{a_0}^{a}\left(\energyt(u)-\frac{2(m-1)}{m-2} \frac{\zeta(u)}{u}\energy_h(u) \right)\, e^{2 \xi(u)}\, du -  \frac{1}{2\mathsf{C}_m} \, \int_{a_0}^{a}\frac{d}{du}\left(e^{2 \xi(u)}\energy_h(u) \right)\, du \\
&=& \frac{1}{2\mathsf{C}_m} \,\left(\energy_h(a_0)- e^{2 \xi(a)}\energy_h(a)\right) + \frac{1}{2\mathsf{C}_m} \,\int_{a_0}^{a}\left(\energyt(u)-\frac{2(m-1)}{m-2} \frac{\zeta(u)}{u}\energy_h(u) \right)\, e^{2 \xi(u)}\, du    \\
&=& \frac{1}{2\mathsf{C}_m} \,\left(\energy_h(a_0)- e^{2 \xi(a)}\left(  \energy_h(a) +\energyttt(a) +\energy_v(a) \right)  \right)\\   
&=& \frac{1}{2\mathsf{C}_m} \,\left(\energy_h(a_0) +\energytt(a,x) - e^{2 \xi(a)}\energy(a,x)  \right)\, .    
\end{eqnarray*}
This implies that
\begin{equation}
2\, \int_0^{s} \beta(u,x)du = 2\, s_1(a) + K\, \left( \frac{e^{2 \xi(a)}}{a^2}-\frac{1}{a_0^2} \right) -\frac{1}{\mathsf{C}_m}\, \widetilde{K}(a,x)
\end{equation}
that is,
\begin{eqnarray*}
\frac{1}{\hubblet^2} + 2  \int_0^s\beta du 
&=&  2\, s_1 -\frac{K}{a_0^2} + \left( \hubble^2 + \frac{K}{a^2} \right) e^{2 \xi} -\frac{1}{\mathsf{C}_m}\, \widetilde{K}\\
&=&  2\, s_1 -\frac{K}{a_0^2} + \frac{\energy}{\mathsf{C}_m} e^{2 \xi} -\frac{1}{\mathsf{C}_m}\, \widetilde{K}\\
&=&  \hubble_0^2 + \frac{1}{\mathsf{C}_m}\left(\energytt-\energytt_0 - \widetilde{K}\right)\, .
\end{eqnarray*}
The rest follows from the theorem \ref{Thm:InequalitiesMaxMin} and the identities  
\begin{equation*}
\frac{1}{\omega} = \max_\K|\hubble_0| \quad \textrm{and} \quad \frac{1}{\Omega} = \min_\K|\hubble_0|  \, . 
\end{equation*}
\qed

When the universe expands from $a$ to $a + \Delta a$, the local $\submetrict$-volume changes $\Delta V$, and an amount $\Delta E$ of energy is transferred from the barotropic-viscous part of the fluid to the homogeneous one, where 
\begin{equation}\label{Eq:TransferEnergyInterpretation}
    \energyt = \lim_{\Delta a \to 0} \,  \lim_{\Delta V  \to 0} \,  \frac{\Delta E}{\Delta a\, \Delta V} \, .
\end{equation}
Even though we have not established the notion of gravitational energy in our model, which would be a huge undertaking in itself, or even an impossible one, $\energy$ and the equation (\ref{Eq:EnergyDensityResidualInequalitiesQ}) express the notion that energy is exchanged within the matter components as the universe expands from a hot and homogeneous distribution of matter to an inhomogeneous configuration. In the process of forming clusters of matter together with huge voids, as matter falls into itself during expansion, much of the "gravitational energy" should be transferred through $\energyt$ to expansion, while the inhomogeneous part $\energytt$, due to the inequality \eqref{Eq:EnergyDensityResidualInequalities}, behaves effectively as a "$e^{-2\xi(a)}$" term, in general.\footnote{If the curvature and the viscosity become dominant in the cosmic dynamics, there might be an extra term attached to the inhomogeneities bounded by "$e^{-2\xi(a)}\widetilde{K}(a,x)$".}  
%
%

In resume, corollary \ref{Thm:EnergyDensity} tells us how the energy density behaves on large scales by singling out the terms responsible for viscosity and the transfer of energy within the matter/radiation components, and also by taming the time evolution of inhomogeneities through the inequality \eqref{Eq:EnergyDensityResidualInequalities}. As we join this result to the idea of a large-scale effective model built on a periodic cosmological spacetime \cite{LGGomes_2024_CQG_1}, we open up new possibilities for facing the standard approach to cosmology.

\subsection{The density contrast}
\label{Sec:DensityContrast}

The density contrast is often the principal measure of inhomogeneity in a cosmological model. We recall that $\langle f \rangle$ is the mean value of a $\Gamma$-periodic function $f$ defined in formula \eqref{Eq:AverageValue}. The following corollary helps us trace its behaviour in our periodic models: 
\begin{cor}\label{Thm:EnergyContrast}
Let be given a solution of Einstein's equations as in corollary \ref{Thm:ExistenceUniquenessHubble}. The density contrast,
\begin{equation*}
\delta 
= \frac{\energy-\langle \energy \rangle }{\langle \energy \rangle }
= \left(\frac{\energytt -\langle \energytt \rangle }{\langle \energy \rangle }\right) e^{-2\xi}\, ,
\end{equation*}
is, according to the notation of corollary \ref{Thm:EnergyDensity}, 
\begin{equation}
\delta(a,x) = \frac{\left(\energytt_0-\langle\energytt_0\rangle\right) + \left(\widetilde{K}-\langle\widetilde{K}\rangle\right) - \mathsf{C}_m\left(\hubble_0^2-\langle\hubble_0^2\rangle\right) + \widetilde{\energytt}}{\left(\energy_h+ \energy_v + \energyttt\right)\, e^{2 \xi} + \langle \energytt \rangle} 
 \, ,
\end{equation}
where $ \widetilde{\energytt}$ is bounded as 
\begin{equation}
|\widetilde{\energytt}| \le \hubble_{max}^2-\hubble_{min}^2 
 \, .
\end{equation}
\end{cor}

We note that the inequalities in the corollaries \ref{Thm:EnergyDensity} and \ref{Thm:EnergyContrast} simplify in the case of intrinsically homogeneous and isotropic spacetimes \cite{LGGomes_2022_CQG_2}, for which the curvature $K$ is constant, and therefore $\widetilde{K}=\langle\widetilde{K}\rangle$.

\section{The effective model arising on large scales}
\label{Sec:EffectiveStandardModel}

\subsection{Considerations on the transfer-energy density in the light of the 
the standard model of cosmology}
\label{Sec:HomogeneousMatterEnergyExchange}

An instructive example is given as we demand the universe to be filled with homogeneous dark energy and cosmic radiation distributions followed by an inhomogeneous distribution of dust. In mathematical terms, we set
\begin{equation*}
m=4 \quad , \quad a_0=1 \quad , \quad  
\quad \textrm{and} \quad  \eqstate=\zeta=0 \, ,
\end{equation*}
so that 
\begin{equation}
e^{2\xi} = a^3
\quad  \textrm{and} \quad 
s= \frac{1}{3} \left(1-a\right) \, .
\end{equation}
We assume the energy density to be composed of dark energy and radiation, which are separately conserved, together with a homogeneous interaction term $\energy_h^{(I)}$ accounting for the energy that the matter falling into itself transfer to the large-scale cosmic dynamics. Hence, the energy density, following corollary \ref{Thm:EnergyDensity}, takes the form 
\begin{equation}
\energy(a,x) = \underbrace{\left(\Omega_\Lambda + \frac{\Omega_r}{a^4}\right)\energy_c + \energy_h^{(I)}(a)}_{\energy_h} + \energyttt(a) + \frac{\energytt (a,x)}{a^3} \, , 
\end{equation}
with $\energy_c=3\, \langle \hubble_0^2 \rangle$ playing the same role as the standard model's critical density \cite{Book_2012_ellis_mac_marteens} of a reference value for the energy density. The relativistic pressure  is
\begin{equation}
\pressure (a,x) = \pressure_h(a) =\left( -\, \Omega_\Lambda + \frac{1}{3}\frac{\Omega_r}{a^4}\right)\energy_c + \pressure_h^{(I)}(a) \, , 
\end{equation}
where $\energy_h^{(I)}$ and $\pressure_h^{(I)}$ satisfy
\begin{equation}\label{Eq:EnergyDensityResidualInequalitiesQExample}
\frac{d\energy_h^{(I)}}{da} + \frac{3}{a}\left(\energy_h^{(I)}+\pressure_h^{(I)} \right)= \energyt \, . 
\end{equation}
We define the dimensionless density parameter for the matter component as $\Omega_m$, with
\begin{equation}
\energytt = \Omega_m\, \energy_c + \delta\energytt \qquad , \qquad \Omega_m = \frac{\langle \energytt_0\rangle}{\energy_c} \, .
\end{equation}
Hence, according to corollary \ref{Thm:EnergyDensity}, the mean value of the energy density becomes
\begin{equation}\label{Eq:LCDMEnergyDensity}
\langle \energy\rangle = \underbrace{\left(\Omega_\Lambda + \frac{\Omega_m}{a^3} +\frac{\Omega_r}{a^4}\right)\energy_c}_{\Lambda \textrm{CDM}}  
+   
\underbrace{\energy_h^{(I)}(a) + \energyttt(a)}_{\textrm{New}} 
+ 
\underbrace{\frac{\langle \delta\energytt\rangle }{a^{3}}}_{\textrm{small}} \, ,
\end{equation}
with $\langle \delta\energytt\rangle$ satisfying $\langle \delta\energytt_0\rangle =0$ and bounded by
\begin{equation}\label{Eq:DeltaInequality}
\mathsf{C}_m\left(\hubble_{min}^2 - \langle\hubble_0^2\rangle\right) 
\le \langle \delta\energytt\rangle \le 
\mathsf{C}_m\left(\hubble_{max}^2 - \langle\hubble_0^2\rangle\right) \, .
\end{equation}

The standard model of cosmology starts from the $\Lambda$CDM term and accesses the $\langle \delta\energytt\rangle/a^{3}$ counterpart through perturbations \cite{Book_2012_ellis_mac_marteens}. However, the interacting part, $\energy_h^{(I)}(a) + \energyttt(a)$, which represents the net energy transferred from the clustering matter to expansion as it falls into its proper gravitational field, is completely new, and cannot be induced from perturbations of the homogeneous model. Furthermore, according to corollary \ref{Thm:EnergyContrast}, the density contrast becomes
\begin{equation}
\delta(a,x) = \frac{\left(\left(\energytt_0-\langle\energytt_0\rangle\right) - \mathsf{C}_m\left(\hubble_0^2-\langle\hubble_0^2\rangle\right) + \widetilde{\energytt}\right)\, a}{\Omega_r + \left(\Omega_m + \langle \delta\energytt \rangle\right) a  +\left(\Omega_\Lambda  + \energy^{(I)} + \energyttt \right)\, a^4} .
\end{equation}
Assuming $\Omega_r \ne 0$ and $a_{min}=0$, this shows that our model "homogenizes" for $a \to 0$ at least as fast as $\delta \sim a$, which implies an earlier homogeneous and isotropic universe.

\subsection{An example for the transfer-energy density}
\label{SubSec:ExampleEnergyDensity}

Let us assume by now that the interaction parameters, represented by the $\energy^{(I)}_h$ and $\pressure^{(I)}_h$, satisfy a barotropic equation of state and that the dust, as it tends to clump together, releases energy for expansion at a steady rate, that is, according to \eqref{Eq:TransferEnergyInterpretation}, 
\begin{equation}
\Omega_E\, \energy_c = \lim_{\Delta a \to 0} \,  \lim_{\Delta V_0  \to 0} \,  \frac{\Delta E}{\Delta a\, \Delta V_0}
\end{equation}
is constant for the limit $\Delta V_0 \to 0$ taken in the co-moving geometry of the observers. These considerations are written as
\begin{equation}
\frac{\energyt}{\energy_c} =  \frac{\Omega_E}{a^3} 
\quad  \textrm{and} \quad  \pressure_h^{(I)} = \eqstate_h\, \energy_h^{(I)} \, .
\end{equation}
The relations above allow us to solve the equation \eqref{Eq:EnergyDensityResidualInequalitiesQExample}, which gives us the energy density
\begin{equation}\label{Eq:EnergyDensityExample1}
\frac{\energy}{\energy_c} = \Omega_\Lambda  - \frac{\alpha\, \Omega_{E}}{a^{2}} + \frac{\Omega_m + \Omega_{E}}{a^{3}} +\frac{\Omega_r}{a^4} + \frac{\Omega_{h}}{a^{3(1+\eqstate_h)}} + \frac{\delta\energytt }{a^{3}}   \, , 
\end{equation}
where $\Omega_h$ is a constant and 
\begin{equation}
\alpha=\frac{3\eqstate_h}{(1+3\eqstate_h)} \, .
\end{equation} 
We observe that, due to the inequality \eqref{Eq:DeltaInequality}, $\langle \delta\energytt \rangle << \energy_c$ is a good approximation at least around $a=1$. In this regime, the effective $\Lambda$CDM-like model is driven by the large-scale energy density
\begin{equation}\label{Eq:EnergyDensityStandardModel}
\frac{\langle\energy\rangle}{\energy_c} \approx \Omega_\Lambda  - \frac{\alpha\, \Omega_{E}}{a^{2}} + \frac{\Omega_m + \Omega_{E}}{a^{3}} +\frac{\Omega_r}{a^4} + \frac{\Omega_{h}}{a^{3(1+\eqstate_h)}}    \, . 
\end{equation}
%

\subsection{A class of exact solutions exemplifying the behaviour of the 
inhomogeneities in the dust regime}
\label{Sec:ClassExactSolutions}

We consider the model investigated in section \ref{SubSec:ExampleEnergyDensity} to be flat ($K(x)=0$) with the additional constraints
\begin{equation}
\Omega_\Lambda=\Omega_r=\Omega_h = 0 
\end{equation}
in the equation \eqref{Eq:EnergyDensityExample1}, for the sake of simplicity. In this case, $\Omega_E$ is obtained by taking the average value of equation \eqref{Eq:EnergyDensityExample1} for $a=1$, that is, 
\begin{equation}
\Omega_E = \frac{1-\Omega_m}{1-\alpha} = (1-\Omega_m)\, (1+3\eqstate_h)\, ,
\end{equation}
for $\langle \energy_0 \rangle = \energy_c$ in the flat case and $\langle \delta\energytt_0\rangle =0$, as shown in eq. \eqref{Eq:DeltaInequality}. The parameter $\beta$ is calculated by direct inspection on the formula \eqref{Eq:BetaFunction}, which gives us 
\begin{equation}\label{Eq:FunctionBetaExample}
\beta =  \frac{\epsilon}{L_1^2}
\quad , \qquad L_1 = \sqrt{\left| \frac{2}{ 3(1-\Omega_m)\,\eqstate_h\,\energy_c}\right|}
\end{equation}
with $\epsilon = \textrm{sgn} \left( (\Omega_m-1)\,\eqstate_h \right)$. We assume matter to be periodically distributed along the Euclidian space $\MS=\R^3$ in cubic boxes of volume $L_0^3$ so that $\hubble$, $\energy$, and $\pressure$ are also periodic. The discrete group of symmetries $\Gamma$ is formed by the translations 
\begin{equation}
    (x,y,z) \mapsto (x+n_1L_0, y+n_2L_0, z+n_3L_0) \, , 
\end{equation}
with $(n_1,n_2, n_3) \in \Z^3$. The quotient manifold is the $3$-torus $\K=\T^3=\R^3/\Gamma$. Theorem \ref{Thm:ExistenceUniqueness} assures the existence and uniqueness of the periodic cosmological model $(a_{min},a_0)\times \R^3$ with the metric (\ref{Eq:MetricHubbleForm}) and initial condition $\hubble(a_0,x)=\hubble_0(x) \ne 0$. 

\begin{figure}[!htb]
    \centering
\includegraphics[scale=0.4]{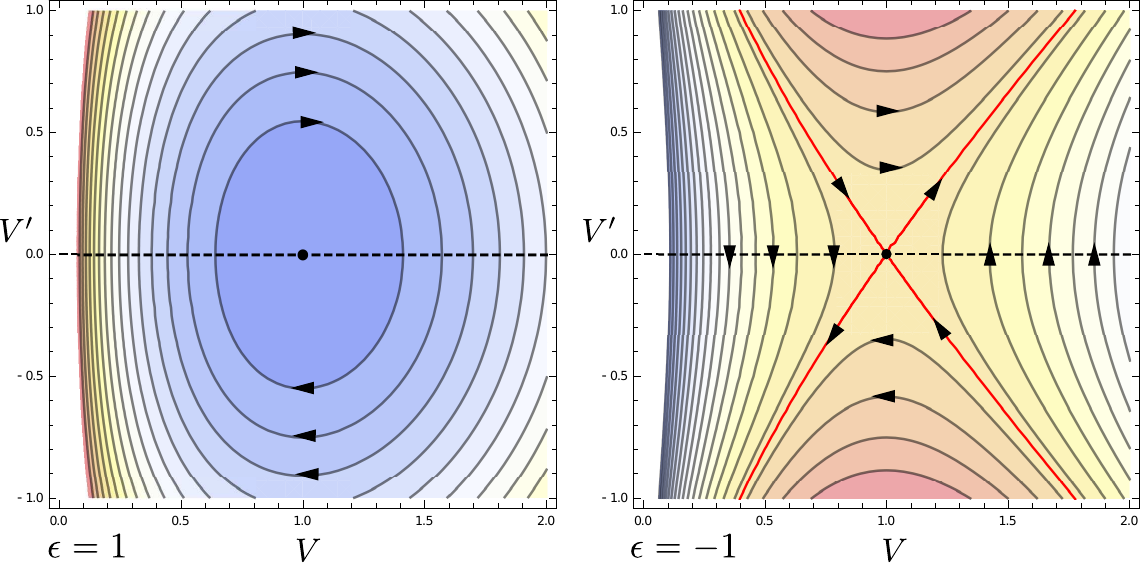}
\caption {Phase portraits for equation (\ref{Eq:PeriodicFlatSeparableEquation}) in the plane $(\hubblev,\hubblev')$. LEFT: all the solutions for $\epsilon =1$ are periodic. RIGHT: no solution but the equilibrium is periodic for $\epsilon =-1$.}
    \label{fig:portraits}
\end{figure}

We specialize a little further by looking for solutions of the equation (\ref{Eq:RestrictedEinsteinFormal}) allowing the energy density to be described by a periodic function of period $T_0$, $\hubblev(\theta)=\hubblev(\theta+T_0)$, in the form 
\begin{equation}\label{Eq:EnergyDensityAnsatz}
\energy = \frac{1}{\hubblev \left(\theta \right)^2}\left( \frac{\omega_m+\omega_E}{a^3}- \frac{\alpha \omega_E}{a^2}\right)
\end{equation}
where
\begin{equation}
\theta = \frac{T_0}{L_0}\left(n_1x+n_2y+n_3z\right)
\end{equation}
and 
\begin{equation}
T_0 = \frac{L_0}{L_1\, \sqrt{n_1^2+n_2^2+n_3^2}}\, .
\end{equation}
Since the equations \eqref{Eq:EnergyDensityExample1} and \eqref{Eq:EnergyDensityAnsatz} must be equal for any values of the parameters, after taking the average values of both at $a=1$, we must have
\begin{equation}
\omega_m = \frac{\Omega_m}{\langle1/\hubblev^2\rangle}
\quad \textrm{and} \quad
\omega_E = \frac{\Omega_E}{\langle1/\hubblev^2\rangle} \, .
\end{equation}
The metric \eqref{Eq:MetricGeneralHomogeneousIsotropic} becomes
\begin{eqnarray}\label{Eq:MetricHubbleFormEuclidian}
 \metric &=& - \, \frac{3\, a \, \langle1/\hubblev^2\rangle \, \hubblev \left(\frac{T_0}{L_0}\left(n_1x+n_2y+n_3z\right)\right)^2}{1+ 3\, \eqstate_h\, (1-\Omega_m)(1-a)}\, da^2  + a^2\, \left(dx^2+dy^2+dz^2\right) \, .
\end{eqnarray}
The function $\hubblev$ is determined by substituting $\energy = 3 \hubble^2 = 3/(a^3 \hubblet^2)$ in the PDE (\ref{Eq:RestrictedEinsteinFormal}). It is governed by the equation
\begin{equation}\label{Eq:PeriodicFlatSeparableEquation}
\frac{d^2 \hubblev}{d\theta^2} +\epsilon\,  \left(\hubblev - \frac{1}{\hubblev}\right) = 0  \,  .
\end{equation}
Fortunately, it is a Hamiltonian dynamical system for the "momentum" $p=V'$, the "position" $q=V$ and the conserved Hamiltonian 
\begin{equation}\label{Eq:Hamiltonian}
\mathcal{H} = \frac{1}{2}((\hubblev')^2+ \epsilon\, \hubblev^2) - \epsilon\,  \ln|\hubblev| \, .
\end{equation}
Therefore, the solutions of \eqref{Eq:PeriodicFlatSeparableEquation} lie on the level sets of \eqref{Eq:Hamiltonian}. They are sketched in Figure \ref{fig:portraits}. We note that equation (\ref{Eq:PeriodicFlatSeparableEquation}) has exactly one equilibrium for $\hubblev>0$, namely $\hubblev=1$, which is a minimum of $\mathcal{H}(\hubblev,\hubblev')$ for $\epsilon=1$ and a saddle point for $\epsilon =-1$. We conclude that the smooth periodic solutions occur only in the first case. From now on, we assume $\epsilon=1$. According to the relation \eqref{Eq:FunctionBetaExample},  this condition is equivalent to 
\begin{equation}
\epsilon = 1 \quad \iff \quad (1-\Omega_m)\,\eqstate_h < 0 \, .
\end{equation}
Figure \ref{fig:portraits} shows the difference between the phase portraits. The symmetry $\hubblev \rightarrow -\hubblev$ of (\ref{Eq:PeriodicFlatSeparableEquation}) shows that we can restrict our analysis to $\hubblev > 0$. 

\begin{figure}[!htb]
    \centering
\includegraphics[scale=0.4]{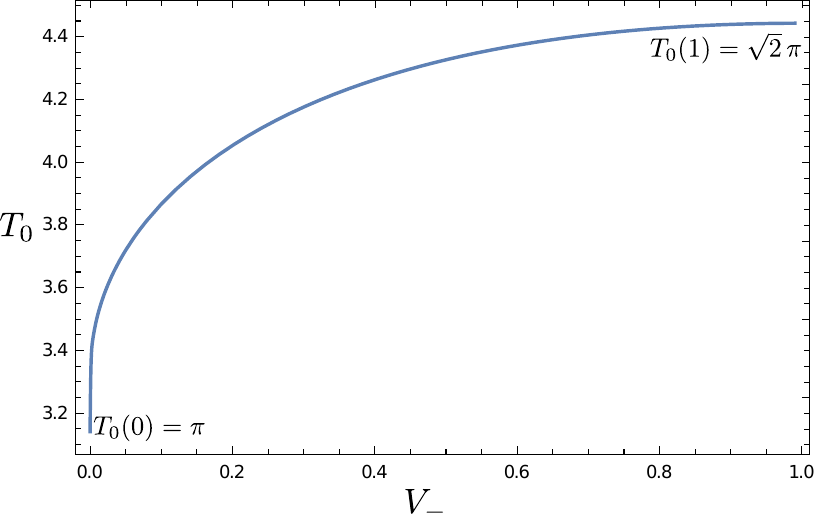}
\caption {Period of the closed orbits as function of $0<\hubblev_{-}<1$, the intersection of the orbit with the horizontal axis.}
    \label{fig:period}
\end{figure}

Each solution $\hubblev(\theta)$ of \eqref{Eq:PeriodicFlatSeparableEquation} with $\epsilon =1$ is characterized by the value $\mathcal{H}_0$ corresponding to its orbit so that it has a unique minimum $0 <\hubblev_{-}<1$ and a unique maximum $\hubblev_{+}>1$ corresponding to the solutions of the algebraic equation
\begin{equation}
\mathcal{H}_0 = \frac{1}{2}\hubblev_{-}^2- \ln\hubblev_{-}= \frac{1}{2}\hubblev_{+}^2- \ln\hubblev_{+} \, .
\end{equation}
The corresponding exact solution is obtained implicitly through the relation 
\begin{equation}
\int  \frac{dV}{\sqrt{\hubblev_{-}^2-\hubblev^2+2 \ln\left(\frac{\hubblev\,\,\,}{\hubblev_{-}} \right)}} = \pm \, \theta \, ,
\end{equation}
where the sign of $\theta$ is suitably chosen. As a consequence, we can compute the period of the complete solution by
\begin{equation}\label{Eq:PeriodExample}
T_0(\hubblev_{-}) = 2 \int_{\hubblev_{-}}^{\hubblev_{+}}\frac{d\hubblev}{\sqrt{\hubblev_{-}^2-\hubblev^2+2 \ln\left(\frac{\hubblev\,\,\,}{\hubblev_{-}} \right)}} \, .
\end{equation}
It is not constant, as Figure \ref{fig:period} shows, and has the lower and upper bounds obtained as 
\begin{equation}
T_{min} = \lim_{\hubblev_{-} \to 0^{+}} T_0(\hubblev_{-}) =  \pi
\qquad \textrm{and}\qquad 
T_{max} = \lim_{\hubblev_{-} \to 1^{-}} T_0(\hubblev_{-}) = \sqrt{2}\, \pi \, .
\end{equation}
The first value was obtained numerically only. The second limit can be calculated analytically with simple manipulations in the integral \eqref{Eq:PeriodExample}.
\begin{figure}[!htb]
    \centering
\includegraphics[scale=0.6]{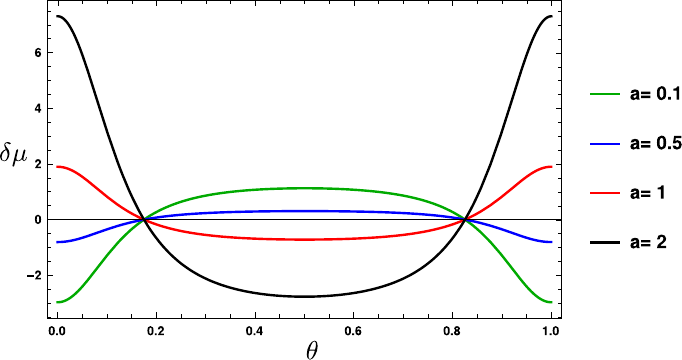}
\caption {An example of the evolution of $\delta\energytt$ with $a$.}
    \label{fig:spatial}
\end{figure}

Finally, the residual energy density in equation \eqref{Eq:EnergyDensityExample1} is obtained as 
\begin{equation}\label{Eq:EnergyDensityDeltaMu}
\delta\energytt = \left(\frac{1}{\langle1/\hubblev^2\rangle\, \hubblev \left(\theta \right)^2}-1\right)\, 
\left( 1+ 3\, \eqstate_h\, (1-\Omega_m)(1-a)\,  \right) \, .
\end{equation}
We first note that $\langle\delta\energytt\rangle=0$, which corroborates the idea that $\langle\delta\energytt\rangle$ is a small term compared to $\energy_c$ in the equation \eqref{Eq:LCDMEnergyDensity}. Moreover, the fact that it is an affine function in the scale factor "a" is compatible with the inequality \eqref{Eq:EnergyDensityResidualInequalities} in the corollary \ref{Thm:EnergyDensity}, for $\energytt = \Omega_m + \delta\energytt$ must be bounded for $0 < a \le 1$, and also tells us this result cannot be extended to $a>a_0$. Moreover, as $a>>1$, matter is aglomerating at the border of the cosmological cells, $x=0$ and $x=L_0$, where its energy density behaves approximately as $\Omega_{Loc}/a^2$, while it is leaving big voids within, near $x=L_0/2$, where its energy density behaves approximately as $-\, \Omega_{Loc}/a^2$, for $\Omega_{Loc} = - 3\, \eqstate_h\, (1-\Omega_m) > 0$ (see figure \eqref{fig:spatial}).

\section{Final remarks}\label{sec:Final Remarks} 
%

%
%
In this manuscript, we have formulated a general mathematical problem arising in Cosmology, which concerns the investigation of spacetimes obeying Einstein's equations constrained by homogeneous and isotropic expansion and carrying the physical interpretation of being spatially homogeneous on ``large scales". We believe this is a two-way connection that can be profitable to both the areas of Mathematical General Relativity and Cosmology. In order to show that, we have formulated a well-posed initial value problem arising from the hypothesis that the energy-momentum tensor is composed of the general homogeneous perfect fluid plus a viscoelastic and inhomogeneous matter. They were analysed in the context of the periodic cosmological models, a mathematical structure suitable for describing homogeneity on ``large scales". We have proved the existence of solutions, their uniqueness and stability under periodic perturbations. The method used to reach such results led us to obtain nontrivial inequalities that, after they were reintroduced to the cosmological context, appear as in equation \eqref{Eq:LCDMEnergyDensity}, where 
\begin{equation}
\energy^{\textrm{eff}} = \energy_h^{(I)}(a) + \energyttt(a)
\end{equation}
is a new term interpreted as the exchange of energy between the homogeneous and inhomogeneous parts of the energy-momentum content, as if the matter is releasing energy to the cosmic expansion whilst it falls into itself by gravitational attraction. Furthermore, we have obtained the density contrast and shown that in the most important context for Cosmology, it demands an early homogeneous period evolving to inhomogeneous distributions of matter, as should be expected. We have also analyzed a class of exact models exemplifying our general assumptions. All those results bring new perspectives for understanding the mathematical foundations of Cosmology as well as prospects for real-world applications. 
%
%
%
%
%

\section*{Acknowledgments}

The authors are thankful for the support from FAPEMIG, project number RED-00133-21. L. R. S.  is also partially supported by FAPEMIG under Grant No.  APQ-02153-23

\section*{Conflict of interest statement }

All authors certify that they have no affiliations with or involvement in any organization or entity with any financial interest or non-financial interest in the subject matter or materials discussed in this manuscript.



%

%
\end{document}